\documentclass[12pt]{article}
\usepackage{graphicx}
\def \b{{\cal B}} 
\def \bea{\begin{eqnarray}}
\def \beq{\begin{equation}}
\def \eea{\end{eqnarray}}
\def \eeq{\end{equation}}
 
\def \ok{\overline{K}^0} 
\def \s{\sqrt{2}} 
\def \st{\sqrt{3}} 
\def \sx{\sqrt{6}}

\begin{document} 
\rightline{EFI 09-04} 
\rightline{TECHNION-PH-2009-05}
\rightline{arXiv:0903.2287} 
\rightline{March 2009} 
\bigskip
\centerline{\bf $D_s$ INCLUSIVE DECAYS} 
\bigskip 
 
\centerline{Michael Gronau\footnote{gronau@physics.technion.ac.il}} 
\centerline{\it Physics Department, Technion -- Israel Institute of Technology} 
\centerline{\it 32000 Haifa, Israel} 
\medskip 
 
\centerline{Jonathan L. Rosner\footnote{rosner@hep.uchicago.edu}} 
\centerline{\it Enrico Fermi Institute and Department of Physics} 
\centerline{\it University of Chicago, 5640 S. Ellis Avenue, Chicago, IL 60637} 
 
\begin{quote}
The availability of branching fractions for a large majority of $D_s$ decays
permits the prediction of inclusive branching fractions.  This is achieved with
the help of a modest amount of input from an isospin statistical model applied
to non-resonant multibody $D_s$ decays.  A systematic uncertainty in these
mostly small branching ratios is estimated by comparing predictions of this
model with those of a model involving quark-antiquark pair production.  The
calculated inclusive branching fractions can be compared with data (for
example, from a large sample of $D_s^+ D_s^{*-} + D_s^{*+} D_s^-$ obtained by
the CLEO Collaboration) and examined for specific final states which can shed
light on strong and weak decay mechanisms.
\end{quote} 

\leftline{PACS numbers: 13.20.Fc, 13.25.Ft, 14.40.Lb, 12.40.Ee} 
\bigskip 
 
\centerline{\bf I.  INTRODUCTION} 
\bigskip 

The mechanisms responsible for decays of hadrons containing heavy quarks
are of interest both as probes of the strong interactions and as sources
of information on the underlying weak processes.  The identification of 
signatures of new physics in such processes often relies on a firm
understanding of long-distance (nonperturbative) effects which can masquerade
as such new physics.

One incompletely understood process in $B$ meson decays is known as
``weak annihilation,'' or WA \cite{WA}.  One wishes to extract the
Cabibbo-Kobayashi-Maskawa (CKM) matrix element $|V_{ub}|$ from charmless
semileptonic $B$ decays.  These constitute only 2\% of all semileptonic
$B$ decays, as $|V_{ub}/V_{cb}|^2 \simeq 1\%$ while phase space favors
$b \to u \ell \nu$ over $b \to c \ell \nu$ by a factor of 2.  Strategies
thus have evolved to extract the small charmless semileptonic fraction.  These
include, for example, the study of leptons with energies $E_\ell$ beyond the
endpoint for $b \to c \ell \nu$.  The WA process can contaminate the endpoint
signal:  a $B^+$ can turn into a soft $I=0$ hadronic system plus a vector
$\bar b u$ which can then annihilate freely into $\ell \nu$.  (Helicity
arguments greatly suppress the annihilation of a pseudoscalar $\bar b u$ into
$\ell \nu$.)

The CLEO \cite{CLEOWA} and BaBar \cite{Aubert:2007tw} Collaborations have
placed upper limits for WA of a few percent of charmless semileptonic $b$
decays, while theoretical estimates \cite{Gambino:2007} lie somewhat lower.
The WA process is supposed to be of order $1/m_Q^3$, where $Q$ is the heavy
quark, so it should be more visible in charm decays.  A suggestion for
probing this process in the semileptonic decay $D_s^+ \to \omega \ell^+
\nu_\ell$ was made in Ref.\ \cite{Gronau:2009mp}.  It was pointed out there
that by comparing the decays $D_s^+ \to \omega \ell^+ \nu_\ell$ and
$D_s \to \phi \ell^+ \nu_\ell$ with the corresponding hadronic decays
$D_s^+ \to \omega \pi^+$ and $D_s^+ \to \phi \pi^+$, one could anticipate
a WA contribution to the branching fraction $\b(\omega \ell^+ \nu_\ell)$ of
order $10^{-3}$, nearly an order of magnitude greater than one would expect
from the process $D_s \to \phi \ell^+ \nu_\ell$ taking account of
$\omega$--$\phi$ mixing.

The hadronic decays of $D_s$ can shed light on the WA process.
The contribution of the weak process $c \bar s \to u \bar d$ to the decay
$D_s \to \omega \pi^+$ is forbidden by G-parity.  Charmed particle decays to
VP final states (V = vector, P = pseudoscalar) dominated by the annihilation
process $c \bar s \to u \bar d$ do not appear to have a consistent description
within flavor SU(3) \cite{Bhattacharya:2008ke}.  In particular, $D_s \to
\omega \pi^+$ is expected to be suppressed, whereas it is seen with
branching fraction $\b(D_s \to \omega \pi^+) = (2.5 \pm 0.9) \times 10^{-3}$,
while the allowed mode $D_s \to \rho^0 \pi^+$ is ``not seen'' \cite{PDG}.

The presence of certain $D_s$ hadronic decay modes containing an $\omega$ thus
could be regarded as evidence for different mechanisms for WA
\cite{Gronau:2009mp}.  The decay $D_s \to \omega \pi^+$, forbidden by ordinary
annihilation, may proceed through pre-radiation of the $\omega$, whether via
violation of the Okubo-Zweig-Iizuka (OZI) rule~\cite{OZI} or rescattering.  For
instance, the $D_s$ can dissociate into two-meson states such as $D^{(*)0}
K^{(*)+}$ and $D^{(*)+} K^{(*)0}$ which rescatter strongly to $(c \bar s)
\omega$ while the virtual $c \bar s$ state decays weakly to $\pi^+$.  On the
other hand, the decay $D_s \to \omega \pi^+ \pi^0$ would be a possible
signature for ordinary WA caused directly by $c \bar s \to u \bar d$,
where the $u \bar d$ current couples to $\omega\pi^+\pi^0$.
In this case the $c \bar s$ system must exchange at least two gluons with
the final state in order to overcome helicity suppression.
On the other hand, if $c \bar s$ emits a $Q=1$ {\it vector} weak current, which
{\it can} couple to $\omega \pi^+$, a state $s \bar s$ is left over, which can
couple to $\eta$.  (The phase space for $\eta'$ is quite limited.)  One would
then expect to see $D_s \to \omega \pi^+ \eta$.

The study of inclusive $\omega$ production in $D_s$ decays is thus of
interest in shedding light on mechanisms of weak decay and their interplay
with long-distance (nonperturbative) physics.  Additionally, the study of
$\eta$ and particuarly $\eta'$ inclusive production is relevant because fits
based on flavor SU(3) have great difficulty in reproducing the large reported
branching fractions $\b(D_s^+ \to \rho^+ \eta) = (13.0 \pm 2.2)\%$ and
$\b(D_s^+ \to \rho^+ \eta') = (12.2 \pm 2.0)\%$ \cite{Bhattacharya:2008ke},
preferring values a factor of 2 and 4 less, respectively.

The CLEO Collaboration has studied the reaction $e^+ e^- \to (D_s^+ D_s^{*-} +
D_s^{*+} D_s^-)$ at $\sqrt{s} = 4.17$ GeV ($\sigma \simeq 1~{\rm nb}^{-1}$)
accumulating a sample of about 600 nb$^{-1}$ \cite{Alexander:2009ux}.  The
dominant decay mode of $D_s^{*\pm}$ is to $\gamma D_s^\pm$.  It then becomes
possible to study a cleanly-identified sample of tagged $D_s$ decays, and thus
to obtain inclusive $D_s$ branching ratios.

In the present paper we undertake a theoretical study of inclusive branching
fractions in $D_s^+$ decays, augmenting the extensive list of known processes
\cite{PDG} with estimates for unobserved modes based on isospin arguments, 
particularly those employing a statistical model \cite{istat,Peshkin:1976kw,QR}.  
The statistical model is described briefly in Sec.\ II.  Systematic
uncertainties in the predictions of this model are obtained by comparison with
predictions of an alternative model involving quark-aniquark pair production
\cite{Eilam:1980fp}. This model is discussed in Sec.\ III.  We then proceed
from known $D_s$ decay modes to others in Sec.\ IV, showing that we thereby
account for the majority of $D_s$ decays and presenting a provisional table of
all $D_s$ branching fractions in Sec.\ V.  These are then employed to construct
a table of inclusive branching fractions in Sec.\ VI, while Sec.\ VII concludes.
\bigskip 

\centerline{\bf II.  STATISTICAL MODEL}
\bigskip

In the absence of information on branching fractions for certain nonresonant 
multibody  modes, we may
use a statistical isospin model to relate them to other known modes.  Such a
model may be constructed by coupling the internal subsystems to isospin
amplitudes in all possible ways and then assuming the reduced amplitudes are
equal in magnitude and incoherent in phase \cite{istat,Peshkin:1976kw,QR}.  We
illustrate this technique with two cases of total $I=I_3=1$, $K \bar K \pi$
and $3 \pi$, using particle orders consistent with those quoted for the
isospin Clebsch-Gordan coefficients in Ref.\ \cite{PDG}.
\bigskip

\noindent
{\bf A.  Example of $(K \bar K \pi)_{I=I_3=1}$}
\medskip

1.  $(K \bar K)\pi$:  We label the reduced amplitudes $A_I^{K \bar K}$ by
$I_{K \bar K} = 0,1$.  Then
\bea
A(K^+ K^- \pi^+)&=&\frac{1}{\s} A^{K \bar K}_0-\frac{1}{2}A^{K \bar K}_1~,\\
A(K^0 \ok \pi^+)&=&-\frac{1}{\s}A^{K \bar K}_0-\frac{1}{2}A^{K \bar K}_1~,\\
A(K^+ \ok \pi^0)&=&\frac{1}{\s}A_1^{K \bar K}~,
\eea
so assuming incoherent and equal amplitudes $A_{0,1}^{K \bar K}$,
\beq \label{eqn:kkbpi}
|A(K^+ K^- \pi^+)|^2:|A(K^0 \ok \pi^+)|^2:|A(K^+ \ok \pi^0)|^2 = 3:3:2~.
\eeq

2.  $(\pi K) \bar K$:  We label the reduced amplitudes $A_I^{\pi K}$
by $I_{\pi K} = 3/2,1/2$.  Then
\bea
A(\pi^+ K^+ K^-) & = &  \frac{\st}{2}A^{\pi K}_{3/2}~,\\
A(\pi^+ K^0 \ok) & = & -\frac{1}{2\st}A^{\pi K}_{3/2}
 + \sqrt{\frac{2}{3}} A^{\pi K}_{1/2}~,\\
A(\pi^0 K^+ \ok) & = & -\frac{1}{\sx}A^{\pi K}_{3/2}
 - \frac{1}{\st} A^{\pi K}_{1/2}~,
\eea
leading again to the ratios (\ref{eqn:kkbpi}) if $A_{1/2,3/2}^{\pi K}$
are equal and incoherent.

3.  $(\pi \bar K) K$:  We label the reduced amplitudes $A_I^{\pi \bar K}$
by $I_{\pi \bar K} = 3/2,1/2$.  Then
\bea
A(\pi^+ K^- K^+) = -\frac{1}{2\st}A^{\pi \bar K}_{3/2}
 + \sqrt{\frac{2}{3}}A^{\pi \bar K}_{1/2}~,\\
A(\pi^+ \ok K^0) = \frac{\st}{2}A^{\pi \bar K}_{3/2}~,\\
A(\pi^0 \ok K^+) = -\frac{1}{\sx}A^{\pi \bar K}_{3/2}
 - \frac{1}{\st}A^{\pi \bar K}_{1/2}~,
\eea
leading again to (\ref{eqn:kkbpi}) if $A_{1/2,3/2}^{\pi \bar K}$
are equal and incoherent.
\bigskip

\noindent
{\bf B.  Example of $(3 \pi)_{I=I_3=1}$}
\medskip

The only couplings to consider are $(\pi \pi) \pi$.  For example, choosing
a particular order,
\bea
A(\pi^+ \pi^+ \pi^-) & = & \sqrt{\frac{3}{5}} A_2~,\\
A(\pi^+ \pi^0 \pi^0) & = & -\sqrt{\frac{3}{20}}A_2 + \frac{1}{2}A_1~,
\eea
so if $A_2$ and $A_1$ are equal and incoherent,
\beq \label{eqn:3pi}
|A(\pi^+ \pi^+ \pi^-)|^2:|A(\pi^+ \pi^0 \pi^0)|^2 = 3:2~.
\eeq
Coupling the pions in a different order one can encounter also an amplitude
with $I_{\pi \pi} = 0$, but the same result is obtained.
\bigskip

{\bf C.  Tables of relative branching fractions}
\medskip

Results for Cabibbo-favored $D_s$ decays are quoted from Ref.\
\cite{Peshkin:1976kw} for $(K \bar K n \pi)_{I=I_3=1}$ final states in
Table \ref{tab:kknpi} and for $(3 \pi)_{I=I_3=1}$ in Table \ref{tab:npi}.
Higher-multiplicity
$D_s$ decays appear to account for a very small fraction of the total
\cite{PDG}.  Results for singly-Cabibbo suppressed $D_s$ decays to $K +
(n \pi)$ can be transcribed from Table I of Ref.\ \cite{QR}, which applies
to a statisical average of $I=1/2$ and $I=3/2$ states for $\bar D^0 \to K +
(n \pi)$ arising from $\bar s \bar u d u$.  Here the $K + (n \pi)$ states arise
from $\bar s \bar d u d$, which is related to $\bar s \bar u d u$ by isospin
reflection.  The results are shown in Table \ref{tab:knpi}.
%
\begin{table}[h]
\caption{Statistical model predictions for charge states in
$(K \bar K n \pi)_{I=I_3=1}$.
\label{tab:kknpi}}
\begin{center}
\begin{tabular}{r c c c c c} \hline \hline
$n(\pi^+) + n(\pi^-)$ & 0 & 1 & \multicolumn{2}{c}{2} & 3 \\
$Q(\bar K)$ &   0  & $-$ or 0 &  $-$ &   0  & $-$ or 0 \\ \hline
$n=0$       &   1  &     --   &  --  &  --  &    --    \\
   1        &  1/4 &    3/8   &  --  &  --  &    --    \\
   2        & 1/10 &   9/40   & 3/20 & 3/10 &    --    \\
   3        & 1/30 &   7/60   & 2/15 & 4/15 &    1/6   \\ \hline \hline
\end{tabular}
\end{center}
\end{table}
%
\begin{table}[h]
\caption{Statistical model predictions for charge states in
$(n \pi)_{I=I_3=1}$.
\label{tab:npi}}
\begin{center}
\begin{tabular}{r c c c} \hline \hline
$n(\pi^+) + n(\pi^-)$ & 1 & 3 & 5 \\ \hline
       $n=1$         & 1 & -- & -- \\
          2          & 1 & -- & -- \\
          3          & 2/5 & 3/5 & -- \\
          4          & 1/5 & 4/5 & -- \\
          5          & 3/35 & 22/35 & 10/35 \\
          6          & 1/28 & 12/28 & 15/28 \\ \hline \hline
\end{tabular}
\end{center}
\end{table}
%
\begin{table}[h]
\caption{Statistical model predictions for charge states in
$K + (n \pi)$ arising from singly-Cabibbo-suppressed $D_s$ decays.  A
statistical average of contributions from $I=1/2$ and $I=3/2$ final states
has been taken as in Table I of Ref.\ \cite{QR}.
\label{tab:knpi}}
\begin{center}
\begin{tabular}{r c c c c} \hline \hline
$n(\pi^+) + n(\pi^-)$ & 0 & 1 & 2 & 3 \\ \hline
       $n=1$         & 1/2 & 1/2 & -- & -- \\
          2          & 3/20 & 8/20 & 9/20 & -- \\
          3          & 3/45 & 9/45 & 21/45 & 12/45 \\ \hline \hline
\end{tabular}
\end{center}
\end{table}

\newpage
\centerline{\bf III. QUARK-ANTIQUARK PAIR PRODUCTION MODEL}
\bigskip
In order to obtain estimates for systematic theoretical uncertainties in 
predictions of the statistical model we present in this section predictions 
of an alternative model involving quark-antiquark ($q \bar q$) pair production.
\bigskip

\noindent{\bf A. Description of the model}
\medskip

In this model final nonresonant states involving several pseudoscalar mesons 
are formed by the production of
$u\bar u$ and $d\bar d$ quark pairs in every possible way, in association 
with a corresponding four-quark process in hadronic $D_s$ decays, and in
association with the $u \bar d$ weak current in semileptonic $\tau$ decays.
One assumes that amplitudes for a given number of quark pairs are equal in
magnitudes and incoherent in phase. For completeness, we describe below quark
diagrams contributing to amplitudes for the three distinct classess studied in
Section II:

\begin{itemize}

\item Cabibbo-favored $D_s\to n\pi$ and $\tau^+\to (n\pi)^+\bar\nu_\tau$. Here 
one fills in $q \bar q$ pairs ($q=u,d$) between the $u \bar d$ pair produced by 
the weak current. 

\item Cabibbo-favored $D_s \to K\bar K\,n\pi$. These decay processes obtain
contributions from two tree amplitudes which are assumed to add incoherently,
a ``color-favored" amplitude ($T$) and a ``color-suppressed" amplitude ($C$).
In the first amplitude ($T$), involving the weak subprocess $c \to (u \bar d)
s$ with a spectator $\bar s$, one fills in with equal probabilities $q \bar q$
pairs between the color-singlet $u \bar d$ pair produced by the weak current 
and between the final $s$ and spectator $\bar s$.  In the second amplitude
($C$), involving the subprocess $c \to (s \bar d) u$ with a spectator $\bar s$,
we fill in with equal probabilities $q \bar q$ pairs between the color-singlet
$s \bar d$ pair in the weak subprocess and between the $u \bar s$ pair.

\item Cabibbo-suppressed $D_s \to K n\pi$.  These decays obtain contributions 
from both color-favored and color-suppressed amplitudes. In the first,
based on the weak subprocess $c \to (u \bar d) d$ with a spectator $\bar s$,
we fill in with equal probabilities $q \bar q$ pairs between the color-singlet
$u \bar d$ pair in the weak subprocess and between the final $d$ and spectator
$\bar s$.  In the color-suppressed amplitude, involving the weak subprocess $c
\to (d \bar d) u$ with a spectator $\bar s$, one fills in with equal
probabilities $q \bar q$ pairs between the color-singlet $d \bar d$ pair in
the weak subprocess and between the final $u$ and spectator $\bar s$.

\end{itemize}
Color-favored ($T$) and color-suppressed ($C$) tree amplitudes, in processes 
involving pions and a $K\bar K$ pair or pions and a kaon, are independent 
quantities assumed to add incoherently in partial decay rates, $\Gamma = |T|^2
+|C|^2$.  Consider two $D_s$ decay processes $D_s\to f_1$ and $D_s\to f_2$
within the same class, $f=K\bar K n\pi$ or $f=K n\pi$ for a given $n$. Denoting
corresponding pairs of amplitudes by $(T_1, C_1)$ and $(T_2, C_2)$ and assuming
for illustration $|C_1|/|C_2| < |T_1|/|T_2|$, one obtains the following lower
and upper bounds on ratios of corresponding partial decay rates or ratios of
branching ratios:
\beq
\frac{|C_1|^2}{|C_2|^2} < \frac{\b_1}{\b_2} = \frac{\Gamma_1}{\Gamma_2} < 
\frac{|T_1|^2}{|T_2|^2}~.
\eeq
Ratios $|T_1|^2/|T_2|^2$ and $|C_1|^2/|C_2|^2$ are calculated in the next
subsection.  They provide lower and upper bounds on ratios of branching ratios.
These bounds, and ratios of branching fractions calculated in this model for
$D_s\to n\pi$ and $\tau^+\to (n\pi)^+\bar\nu_\tau$,
will be compared in Sections IV and V with predictions of the statistical isospin 
model in order to estimate systematic theoretical errors in these predictions.
\bigskip

\noindent{\bf B. Tables of relative branching fractions}
\medskip

Results for relative branching fractions obtained in the $q \bar q$ pair 
production model are shown in Tables \ref{tab:npiPop}, \ref{tab:kknpiPopT}, and
\ref{tab:knpiPopK}.  The upper and lower parts of the last two tables, denoted
$T$ and $C$ respectively, correspond to color-favored and color-suppressed
amplitudes. 
\begin{table}[h]
\caption{Predictions of the $q \bar q$ pair production model for charge
states in $(n \pi)_{I=I_3=1}$.
\label{tab:npiPop}}
\begin{center}
\begin{tabular}{r c c c} \hline \hline
$n(\pi^+) + n(\pi^-)$ & 1 & 3 & 5 \\ \hline
       $n=1$         & 1 & -- & -- \\
          2          & 1 & -- & -- \\
          3          & 3/7 & 4/7 & -- \\
          4          & 1/5 & 4/5 & -- \\
          5          & 5/61 & 40/61 & 16/61 \\ \hline \hline
\end{tabular}
\end{center}
\end{table}
\begin{table}[h]
\caption{Predictions of the $q \bar q$ pair production model for charge 
states in $(K \bar K n \pi)_{I=I_3=1}$, for
color-favored ($T$) and color-suppressed ($C$) amplitudes.
\label{tab:kknpiPopT}}
\begin{center}
\begin{tabular}{r c c c c c} \hline \hline
$n(\pi^+) + n(\pi^-)$ & 0 & 1 & \multicolumn{2}{c}{2} & 3 \\
$Q(\bar K)$ &   0  & $-$ or 0 &  $-$ &   0  & $-$ or 0 \\ \hline
$T$~~~~~~~~~~$n=0$       &   1  &     --   &  --  &  --  &    --    \\
   1        &  0 &    1/2   &  --  &  --  &    --    \\
   2        & 0 &   3/10   & 1/5 & 1/5 &    --    \\
   3        & 0 &   3/22   & 2/11 & 2/11 &  2/11   \\ \hline 
$C$~~~~~~~~~~$n=0$       &   1  &     --   &  --  &  --  &    --    \\
   1        &  1/3 &    1/3   &  --  &  --  &    --    \\
   2        & 1/9 &   2/9   & 4/27 & 8/27 &    --    \\
   3        & 1/27 &   1/9   & 4/27 & 8/27 &   4/27   \\ \hline \hline
\end{tabular}
\end{center}
\end{table}
\begin{table}[h]
\caption{Predictions of the $ q \bar q$ pair production model for charge 
states in $K + (n \pi)$, for
color-favored ($T$) and color-suppressed ($C$) amplitudes.
\label{tab:knpiPopK}}
\begin{center}
\begin{tabular}{r c c c c} \hline \hline
$n(\pi^+) + n(\pi^-)$ & 0 & 1 & 2 & 3 \\ \hline
$T$~~~~~~~~~~$n=1$         & 0 & 1 & -- & -- \\
          2          & 0 & 3/5 & 2/5 & -- \\
          3          & 0 & 6/22 & 4/11 & 4/11 \\ \hline 
$C$~~~~~~~~~~$n=1$         & 1 & 0 & -- & -- \\
          2          & 1/4 & 1/4 & 1/2 & -- \\
          3          & 3/37 & 6/37 & 20/37 & 8/37 \\ \hline \hline
\end{tabular}
\end{center}
\end{table}
\bigskip

\centerline{\bf IV.  FROM KNOWN $D_s$ DECAYS TO OTHER MODES} 
\bigskip

\noindent{\bf A.  Hadronic modes from $D_s^+ \to \tau^+ \nu_\tau$}
\medskip

As the branching ratio $\b(D_s \to \tau \nu_\tau) = (6.6 \pm 0.6)\%$
\cite{PDG,newtau}, and $\tau$ has important decays to hadrons, we must
take account of such decays.  The known $\tau^+$ branching ratios to hadronic
final states are summarized in Table \ref{tab:tau}, with values in brackets
inferred by applying the statistical isospin model to $\b(\tau^+ \to (5 \pi)^+
\bar \nu_\tau) = (7.6 \pm 0.5) \times10^{-3}$ \cite{PDG}.  The second error in
each of the three values of $\b_f$ corresponding to $f^+=(5\pi)^+$ is a
systematic error obtained by comparing predictions of this model with those of
the $q \bar q$ pair production model.  Also shown are inclusive
branching ratios of $\tau^+$ decays to $\pi^+$, $\pi^0$, $\pi^-$, $K^+$, and
$K^0$.  These are multiplied by $\b(D_s \to \tau \nu_\tau)$ to obtain
contributions to inclusive $D_s$ decays.  One sees that the contributions,
particularly to secondary $\pi^+$ and $\pi^0$, are significant.
\bigskip

\begin{table}
\caption{Branching fractions $\b_f \equiv \b(\tau^+ \to f^+ \bar \nu_\tau)$
and contributions to inclusive pion and kaon production.  Modes with branching
fractions less than $10^{-4}$ are omitted.  Brackets denote modes with values
obtained from $\b(\tau^+ \to (5 \pi)^+ \bar \nu_\tau) = (7.6 \pm 0.5) \times
10^{-3}$ \cite{PDG} and the statistical model coefficients of Table
\ref{tab:npi}.  The last line is calculated with $\b(D_s) \equiv \b(D_s \to
\tau^+\nu_\tau)=(6.6 \pm 0.6)\%$ \cite{PDG}.  Contributions to inclusive $\ok$
and $K^-$ production are found to be $\b(\tau^+ \to \ok X \bar \nu_\tau)
= (0.52 \pm 0.05)\%$, $\b(\tau^+ \to K^- X \bar \nu_\tau) = (0.140 \pm 0.005)
\%$, or $0.03$ and $0.01$ when multiplied by $\b(D_s \to \tau \nu_\tau)$.
Columns 3--7 have been rounded off for convenience.
\label{tab:tau}}
{\small
\begin{tabular}{c c c c c c c} \hline \hline
$f^+$ & $\b_f$ (\%) & $\b(\pi^+)(\%)$ & $\b(\pi^0)(\%)$ & $\b(\pi^-)(\%)$ &
 $\b(K^+)(\%)$ & $\b(K^0)(\%)$ \\ \hline
$\pi^+$ & 10.91$\pm$0.07 & 10.91$\pm$0.07 & 0 & 0 & 0 & 0 \\
$\pi^+ \pi^0$ & 25.52$\pm$0.10 & 25.52$\pm$0.10 & 25.52$\pm$0.10 & 0 & 0 & 0 \\
$\pi^+ 2\pi^0$ & 9.27$\pm$0.12 & 9.27$\pm$0.12 & 18.54$\pm$0.24 & 0 & 0 & 0 \\
$2\pi^+ \pi^-$ & 9.03$\pm$0.06 & 18.06$\pm$0.12 & 0 & 9.03$\pm$0.06 & 0 & 0 \\
$2\pi^+ \pi^- \pi^0$ & 4.48$\pm$0.06 & 8.96$\pm$0.12 & 4.48$\pm$0.06 &
 4.48$\pm$0.06 & 0 & 0 \\
$\pi^+ 3\pi^0$ & 1.04$\pm$0.07 & 1.04$\pm$0.07 & 3.12$\pm$0.21 & 0 & 0 & 0 \\
$K^+$   & 0.695$\pm$0.023 & 0 & 0 & 0 & 0.70$\pm$0.02 & 0 \\
$K^0 \pi^+$ & 0.84$\pm$0.04 & 0.84$\pm$0.04 & 0 & 0 & 0 & 0.84$\pm$0.04 \\
$K^+ \pi^0$ & 0.428$\pm$0.015 & 0 & 0.43$\pm$0.02 & 0 & 0.43$\pm$0.02 &
 0 \\
$K^+ 2\pi^0$ & 0.063$\pm$0.023 & 0 & 0.13$\pm$0.05 & 0 & 0.06$\pm$0.02 &
 0 \\
 $K^+ \pi^+ \pi^-$ & 0.287$\pm$0.016 & 0.29$\pm$0.02 & 0 & 0.29$\pm$0.02 &
0.29$\pm$0.02 & 0 \\
$K^0 \pi^+ \pi^0$ & 0.39$\pm$0.04 & 0.39$\pm$0.04 & 0.39$\pm$0.04 & 0 & 0 &
 0.39$\pm$0.04 \\
$K^+ \pi^+ \pi^- \pi^0$ & 0.081$\pm$ 0.012 & 0.08$\pm$ 0.01 &
 0.08$\pm$ 0.01 & 0.08$\pm$ 0.01 & 0.08$\pm$ 0.01 & 0 \\
$K^+ 3\pi^0$ & 0.047$\pm$0.021 & 0 & 0.14$\pm$0.06 & 0 & 0.05$\pm$0.02 &
0 \\
$K^0 \pi^+ 2\pi^0$ & 0.026$\pm$0.024 & 0.03$\pm$0.02 & 0.05$\pm$0.05 &
0 & 0 & 0.03$\pm$0.02 \\
$K^+ \ok$ & 0.158$\pm$0.016 & 0 & 0 & 0 & 0.16$\pm$0.02 & 0 \\
$K^+ \ok \pi^0$ & 0.158$\pm$0.020 & 0 & 0.16$\pm$0.02 & 0 & 0.16$\pm$0.02 &
 0 \\
$K^0 \ok \pi^+$ & 0.17$\pm$0.04 & 0.17$\pm$0.04 & 0 & 0 & 0 & 0.17$\pm$0.04 \\
$K^+ K^- \pi^+$ & 0.140$\pm$0.005 & 0.14$\pm$0.01 & 0 & 0 & 0.14$\pm$0.01
 & 0 \\
$K^0 \ok \pi^+ \pi^0$ & 0.031$\pm$0.023 & 0.03$\pm$0.02 & 0.03$\pm$0.02 &
 0 & 0 & 0.03$\pm$0.02 \\
$[3\pi^+ 2\pi^-]$ & 0.217$\pm$0.014$\pm$0.018 & 0.65$\pm$0.07 & 0 &
 0.43$\pm$0.05 & 0 & 0 \\
$[2\pi^+ \pi^- 2\pi^0]$ & 0.478$\pm$0.031$\pm$0.020 & 0.96$\pm$0.07 &
 0.96$\pm$0.07 & 0.48$\pm$0.04 & 0 & 0 \\
$[\pi^+ 4 \pi^0]$ & 0.065$\pm$0.004$\pm$0.003 & 0.07 & 0.26$\pm$0.02 & 0 &
 0 & 0 \\ \hline
Totals: & & 77.40$\pm$0.28 & 54.28$\pm$0.36 & 14.79$\pm$0.11 & 2.06$\pm$0.05 &
 1.46$\pm$0.08 \\
$\times \b(D_s)$ & & 5.11$\pm$0.46 & 3.58$\pm$0.33 & 0.98$\pm$0.09
 & 0.14$\pm$ 0.01 & 0.10$\pm$0.01 \\
\hline \hline
\end{tabular}
}
\end{table}

\newpage
\noindent{\bf B. Secondary particles from $\eta$, $\eta'$, $\phi$, and
$\omega$}
\medskip

The decays of the $\eta$ contribute to secondary pions and photons.  In
Table \ref{tab:eta} we note these contributions for pions.  Secondary
photons will not be considered here.
\medskip

\begin{table}
\caption{Branching fractions in $\eta$ decays \cite{PDG} and contributions to
inclusive pions.  Branching fractions less than 1\% are omitted.
\label{tab:eta}}
\begin{center}
\begin{tabular}{c c c c c} \hline \hline
$\eta$ mode & $\b(\%)$ & $\b(\pi^+)(\%)$ & $\b(\pi^0)(\%)$ & $\b(\pi^-)(\%)$ \\
 \hline
$\gamma \gamma$ & 39.31$\pm$0.20 & 0 & 0 & 0 \\
$3 \pi^0$ & 32.56$\pm$0.20 & 0 & 97.68$\pm$0.60 & 0 \\
$\pi^+ \pi^- \pi^0$ & 22.73$\pm$0.28 & 22.73$\pm$0.28 & 22.73$\pm$0.28 &
 22.73$\pm$0.28 \\
$\pi^+ \pi^- \gamma$ & 4.60$\pm$0.16 & 4.60$\pm$0.16 & 0 & 4.60$\pm$0.16 \\
\hline
Total & & 27.33$\pm$0.32 & 120.41$\pm$0.66 & 27.33$\pm$0.32 \\ \hline \hline
\end{tabular}
\end{center}
\end{table}

The decays of the $\eta'$ contribute to a number of inclusive final states.
These are summarized in Table \ref{tab:etap} for the main $\eta'$ decay
modes (those with branching fractions greater than 1\%).
\medskip

\begin{table}
\caption{Branching fractions in $\eta'$ decays \cite{PDG} and contributions to
inclusive pions, $\eta$, and $\omega$.  In calculating pion production we
use the branching fractions of Table \ref{tab:eta} for $\eta$ and note that
$\b(\omega \to \pi^+ \pi^- \pi^0) = (89.2\pm0.7)\%$, $\b(\omega \to \pi^0
\gamma) = (8.92 \pm 0.24)\%$, $\b(\omega \to \pi^+ \pi^-) =
(1.53^{+0.13}_{-0.11})\%$ (see Table XI).
\label{tab:etap}}
\begin{center}
\begin{tabular}{c c c c c c c} \hline \hline
$\eta'$ mode & $\b(\%)$ & $\b(\pi^+)(\%)$ & $\b(\pi^0)(\%)$ & $\b(\pi^-)(\%)$ &
 $\b(\eta)(\%)$ & $\b(\omega)(\%)$ \\ \hline
$\pi^+ \pi^- \eta$ & 44.6$\pm$1.4 & 56.8$\pm$1.8 & 53.7$\pm$1.7 & 56.8$\pm$1.8
 & 44.6$\pm$1.4 & 0 \\
$\pi^+ \pi^- \gamma$ & 29.4$\pm$0.9 & 29.4$\pm$0.9 & 0 & 29.4$\pm$0.9 &
 0 & 0 \\
$\pi^0 \pi^0 \eta$ & 20.7$\pm$1.2 & 5.66$\pm$0.33 & 66.3$\pm$3.8 &
 5.66$\pm$0.33 & 20.7$\pm$1.2 & 0 \\
$\omega \gamma$ & 3.02$\pm$0.31 & 2.74$\pm$0.28 & 2.96$\pm$0.30 & 2.74$\pm$0.28
 & 0 & 3.02$\pm$0.31 \\ \hline
Total & & 94.6$\pm$2.1 & 123.0$\pm$4.2 & 94.6$\pm$2.1 & 65.3$\pm$1.8 &
 3.02$\pm$0.31 \\
\hline \hline
\end{tabular}
\end{center}
\end{table}

The $\phi$ is produced copiously in $D_s$ decays.  Its decays to $K^+  K^-$
and $K^0 \ok$ occur with different branching fractions because of differences
in the limited phase space.  Moreover, it has non-negligible branching
fraction into $\pi^+ \pi^- \pi^0$ and $\eta \gamma$.  In Table \ref{tab:phi} we
collect these branching fractions and their implications for inclusive
signals.
\medskip

\begin{table}
\caption{Branching fractions of $\phi$ \cite{PDG} and their implications for
inclusive pion production.  In addition the branching fractions for kaon
and $\eta$ production in $\phi$ decay are $\b(K^+) = \b(K^-) = (49.2 \pm
0.6)\%$, $\b(K^0) = \b(\ok) = (34.2\pm 0.5)\%$, and $\b(\eta) = (1.30 \pm
0.03)\%$.
\label{tab:phi}}
\begin{center}
\begin{tabular}{c c c c c} \hline \hline
$\phi$ mode & $\b(\%)$ & $\b(\pi^+)$(\%) & $\b(\pi^0)$(\%) & $\b(\pi^-)$(\%) \\
 \hline
$K^+ K^-$ & 49.2$\pm$0.6 & 0 & 0 & 0 \\
$K^0 \ok$ & 34.0$\pm$0.5 & 0 & 0 & 0 \\
$\pi^+ \pi^- \pi^0$ & 15.25$\pm$0.35 & 15.25$\pm$0.35 & 15.25$\pm$0.35 &
 15.25$\pm$0.35 \\
$\eta \gamma$ & 1.304$\pm$0.025 & 0.36$\pm$ 0.01 & 1.57$\pm$0.03 &
 0.36$\pm$ 0.01 \\ \hline
Total & & 15.61$\pm$0.35 & 16.82$\pm$0.35 & 15.61$\pm$0.35 \\ \hline \hline
\end{tabular}
\end{center}
\end{table}

Although few $D_s$ modes involve $\omega$, we consider the main modes of
$\omega$ for completeness in Table \ref{tab:om}.

\begin{table}
\caption{Branching fractions in $\omega$ decays \cite{PDG} and contributions
to inclusive pions.
\label{tab:om}}
\begin{center}
\begin{tabular}{c c c c c} \hline \hline
$\omega$ mode & $\b(\%)$ & $\b(\pi^+)(\%)$ & $\b(\pi^0)(\%)$ & $\b(\pi^-)(\%)$
 \\ \hline
$\pi^+\pi^-\pi^0$ & 89.2$\pm$0.7 & 89.2$\pm$0.7 & 89.2$\pm$0.7 & 89.2$\pm$0.7\\
$\pi^0\gamma$ & 8.92$\pm$0.24 & 0 & 8.92$\pm$0.24 & 0 \\
$\pi^+\pi^-$ & $1.53^{+0.11}_{-0.13}$ & $1.53^{+0.11}_{-0.13}$ & 0 &
 $1.53^{+0.11}_{-0.13}$ \\ \hline
Total & & 90.7$\pm$0.7 & 98.1$\pm$0.7 & 90.7$\pm$0.7 \\ \hline \hline
\end{tabular}
\end{center}
\end{table}
\bigskip

\noindent
{\bf C.  Treatment of neutral kaons}
\medskip

Often branching fractions are quoted for decay of a zero-strangeness system
into $X K_S$, where $X$ has a known strangeness.  Thus if $S(X) = 1$, we
infer $\b(X \ok) = 2 \b(X K_S)$, $\b(X K^0) = 0$, while if $S(X) = -1$, we
infer $\b(X K^0) = 2 \b(X K_S)$, $\b(X \ok) = 0$.  Similarly, if a state
of known strangeness contains a single $K_S$ we double the branching fraction
and quote it for the appropriate $K^0$ or $\ok$.  Bracketed states in the table
of branching fractions will indicate modes for which this has been done.
\bigskip

\noindent
{\bf D.  $K \bar K n\pi$ final states}
\medskip

We use the fact that $D_s \to K \bar K \pi$ is dominated by quasi-two-body
$\phi \pi^+$, $\bar K^{*0} K^+$, and $\ok K^{*+}$ final states to calculate
the relative abundances of three charge states, using a statistical model
for the left-over (presumably non-resonant or broad S-wave) component.  We
start with branching ratios quoted in Ref.\ \cite{PDG}: $\b(D_s \to \phi
\pi^+) = (4.38 \pm 0.35)\%$, $\b(D_s \to \bar K^{*0} K^+) = (3.9 \pm 0.6)\%$
(inferred from the quoted value of $(2.6 \pm 0.4)\%$ with $K^{*0}$ decaying
to $K^+ \pi^-$), and $\b(D_s \to \ok K^{*+}) = (5.3 \pm 1.2)\%$.  We
summarize in Table \ref{tab:kkpi} the contributions of each subsystem to
the three $K \bar K \pi$ charge states, as well as to $2 \pi^+ \pi^0 \pi^-$
arising from $\phi \to \pi^+ \pi^- \pi^0$.  Here we have used the branching
fractions for $\phi$ in Table \ref{tab:phi}, and have assumed $K^*$ decays
are governed by isospin Clebsch-Gordan coefficients. 

\begin{table}
\caption{Contributions of quasi-two-body $D_s$ decays to $K \bar K \pi$
and $2 \pi^+ \pi^- \pi^0$ charge states, in percent branching ratio from $D_s$.
\label{tab:kkpi}}
\begin{center}
\begin{tabular}{c c c c c} \hline \hline
Mode & $K^+ K^- \pi^+$ & $K^0 \ok \pi^+$ & $K^+ \ok \pi^0$ & $2 \pi^+ \pi^-
\pi^0$ \\ \hline
$\phi \pi^+$ & 2.15$\pm$0.17 & 1.49$\pm$0.12 & 0 & 0.67$\pm$0.05 \\
$\bar K^{*0} K^+$ & 2.6$\pm$0.4 & 0 & 1.3$\pm$0.2 & 0 \\
$\ok K^{*+}$ & 0 & 3.53$\pm$0.80 & 1.77$\pm$0.40 & 0 \\
Subtotal & 4.76$\pm$0.43 & 5.02$\pm$0.81 & 3.07$\pm$0.45 & 0.67$\pm$0.05 \\
``NR'' & 0.74$\pm$0.51 & 0.74$\pm$0.51 & 0.49$\pm$0.34$^{+0.25}_{-0.49}$ & 0 \\
Total & 5.50$\pm$0.28 & 5.76$\pm$0.96 & 3.56$\pm$0.67 & 0.67$\pm$0.05 \\
\hline \hline 
\end{tabular}
\end{center}
\end{table}

The subtotals represent the sums of quasi-two-body contributions to different
charge states.  As $\b(D_s \to K^+ K^- \pi^+) = (5.50 \pm 0.28)\%$ is measured,
we add a nonresonant (``NR'') or broad S-wave background of $(0.74 \pm 0.51)\%$
to reach this total.  The inferred branching ratios $\b(D_s \to K^0 \ok \pi^+)
= (5.76 \pm 0.96)\%$ and $\b(D_s \to K^+ \ok \pi^0) = (3.56 \pm 0.67)\%$ 
(where the error includes a small contribution from a systematic model
uncertainty) are not very different from the respective predictions of $(5.50
\pm 0.28)\%$ and $(3.67 \pm 0.19)\%$ of a pure statistical model.
\medskip

Observed quasi-two-body decays of $D_s$ contributing to $K \bar K 2 \pi$
final states are to $\phi (\to K^+ K^-) \rho^+$, with branching fraction
($4.0^{+1.1}_{-1.2})\%$, and $K^{*+} \bar K^{*0}$, with branching fraction
$(7.0 \pm 2.5)\%$.  As seen in Table \ref{tab:kk2pi}, these two contributions
almost completely saturate the final state $K^+ K^- \pi^+ \pi^0$, leaving a
very small ``NR'' amount to be treated model-dependently.

\begin{table}
\caption{Contributions of quasi-two-body $D_s$ decays to $K \bar K 2 \pi$
charge states, in percent branching ratio from $D_s$.  Not shown:  A
contribution of $D_s \to \phi \rho^+$ to the $2 \pi^+ \pi^- 2 \pi^0$ final
state with branching fraction $(4.0^{+1.1}_{-1.2})\% \cdot (15.25 \pm 0.35)\%/
(49.2 \pm 0.6)\% = (1.24 \pm 0.36)\%$.  Totals with smaller errors are
observed branching fractions \cite{PDG} and are used instead of model
estimates.  (Branching fractions to $K^+ \ok \pi^+ \pi^-$ and $K^0 K^- 2\pi^+$
are inferred by doubling reported branching fractions involving $K_S$.)
\label{tab:kk2pi}}
\begin{center}
\begin{tabular}{c c c c c c} \hline \hline
Mode & $K^+ K^- \pi^+ \pi^0$ & $K^0 \ok \pi^+ \pi^0$ & $K^+ \ok \pi^+ \pi^-$ &
 $K^0 K^- \pi^+ \pi^+$ & $K^+ \ok \pi^0 \pi^0$ \\ \hline
$\phi \rho^+$ & 4.0$^{+1.1}_{-1.2}$ & 2.76$\pm$0.79 & 0 & 0 & 0 \\
$K^{*+} \bar K^{*0}$ & 1.56$\pm$0.56& 1.56$\pm$0.56 & 0 & 3.11$\pm$1.11 &
 0.78$\pm$0.28 \\
Subtotal & 5.56$\pm$1.28 & 4.32$\pm$0.97 & 0 & 3.11$\pm$1.11 & 0.78$\pm$0.28 \\
``NR'' & 0.04$\pm$1.37 & 0.04$\pm$1.37 & 0.05$\pm$1.83$\pm$0.02 &
 0.03$\pm$0.91 & 0.02$\pm$0.61 \\
Total & 5.6$\pm$0.5 & 4.36$\pm$1.68 & 1.92$\pm$0.26 & 3.28$\pm$0.24 &
 0.80$\pm$0.67 \\ \hline \hline
\end{tabular}
\end{center}
\end{table}
\medskip

The only subsystem for which one has reliable information in $D_s \to K \bar K
3 \pi$ is $\phi 2\pi^+ \pi^-$, with quoted branching fraction $\b(D_s \to
\phi (\to K^+ K^-) 2\pi^+ \pi^-) = (0.59 \pm 0.11)\%$.  Using the $\phi$
branching fractions in Table \ref{tab:phi}, one infers
\bea \nonumber
\b(D_s \to \phi 2\pi^+ \pi^-) & = &(1.20 \pm 0.22)\%~,\\
\b(D_s \to \phi (\to K^+ K^-) 2\pi^+ \pi^-) & = & (0.59 \pm 0.11)\%~,
\nonumber \\
\b(D_s \to \phi (\to K^0 \ok) 2\pi^+ \pi^-) & = & (0.41 \pm 0.08)\%~,
\nonumber \\
\b(D_s \to \phi (\to \pi^+ \pi^- \pi^0) 2\pi^+ \pi^-) & = & (0.18\pm0.03)\%~.
\eea
The $\pi^+ \pi^0 \pi^0$ system accompanies the $\phi$ 2/3 and 3/4 as frequently as 
$2 \pi^+ \pi^-$ in the statistical model and the $q\bar q$ production model, respectively
(see Tables \ref{tab:npi} and \ref{tab:npiPop}).  Thus one gets the corresponding 
predictions
\bea \nonumber
\b(D_s \to \phi \pi^+ 2\pi^0) & = &(0.80 \pm 0.15\pm 0.10)\%~,\\
\b(D_s \to \phi (\to K^+ K^-) \pi^+ 2\pi^0) & = & (0.39 \pm 0.07\pm 0.05)\%~,
\nonumber \\
\b(D_s \to \phi (\to K^0 \ok) \pi^+ 2\pi^0) & = & (0.27 \pm 0.05\pm 0.04)\%~,
\nonumber \\
\b(D_s \to \phi (\to \pi^+ \pi^- \pi^0) \pi^+ 2\pi^0) & = & (0.12\pm0.02\pm 0.02)\%~,
\eea
where the second error represents the difference between predictions of the two
models.
The only $D_s \to K \bar K 3 \pi$ modes for which branching fractions have
been measured are $D_s^+ \to K^+ K^- 2\pi^+ \pi^-$, with $\b = (0.88
\pm 0.16)\%$, and $D_s \to K_S K_S 2\pi^+ \pi^-$, with $\b = (8.4 \pm 3.5)
\times 10^{-4}$.  It may be risky to use the latter to extrapolate to
the $K^0 \ok 2\pi^+ \pi^-$ mode if the kaon pair favors one CP eigenstate over
another.  Consequently, we shall use the difference between $\b(D_s \to
K^+ K^- 2\pi^+ \pi^-)$ and its contribution from the $\phi$ to estimate
the amount of the $K \bar K 3 \pi$ final state to be treated model-dependently.
The results are shown in Table \ref{tab:kk3pi}, where symmetric systematic 
errors are quoted using the maximum of positive and negative errors.

\begin{table}
\caption{Contributions in percent of $\phi(3 \pi)$ and model contributions
to charge states in $D_s \to K \bar K 3 \pi$. The latter contributions, based 
on the statistical model, contain a statistical error and systematic error
obtained by comparison with the $q\bar q$ production model. 
\label{tab:kk3pi}}
\begin{center}
\begin{tabular}{c c c c}  \hline \hline
State & $\phi(3 \pi)$ & Model & Total \\ \hline
$K^+ K^- 2 \pi^+ \pi^-$ & 0.59$\pm$0.11 & 0.29$\pm$0.19 & 0.88$\pm$0.16 \\
$K^0 \ok 2 \pi^+ \pi^-$ & 0.41$\pm$0.08 & 0.29$\pm$0.19 & 0.70$\pm$0.21 \\
$K^+ \ok \pi^+ \pi^- \pi^0$& 0 & 0.46$\pm$0.30$\pm$0.17 & 0.46$\pm$0.34 \\
$K^0 K^- 2\pi^+ \pi^0$ & 0 & 0.23$\pm$0.15$\pm$0.06 & 0.23$\pm$0.16 \\
$K^+ K^- \pi^+ 2\pi^0$ & 0.39$\pm$0.09 & 0.20$\pm$0.13$\pm$0.02 & 0.59$\pm$0.16 \\
$K^0 \ok \pi^+ 2\pi^0$ & 0.27$\pm$0.06 & 0.20$\pm$0.13$\pm$0.02 & 0.47$\pm$0.15 \\
$K^+ \ok 3\pi^0$ & 0 & 0.06$\pm$0.04$\pm$0.06 & 0.06$\pm$0.06 \\ \hline \hline
\end{tabular}
\end{center}
\end{table}

\bigskip

\noindent
{\bf E.  Multiple pions from $\phi (n \pi)$ final states}
\medskip

We have already calculated branching fractions for $D_s$ to decay to multipion
states via the mode $\phi \to \pi^+ \pi^- \pi^0$, but summarize the results
here.

(1) The state $2\pi^+ \pi^- \pi^0$ receives a contribution from
$D_s \to \phi \pi^+$, resulting in $\b(D_s^+ \to 2\pi^+ \pi^- \pi^0) =
(4.38 \pm 0.35)\% \cdot (15.25 \pm 0.35)\% = (0.67 \pm 0.06)\%$.

(2) The state $2\pi^+ \pi^- 2\pi^0$ receives a contribution from $D_s \to \phi
\rho^+$, resulting in $\b(D_s^+ \to 2\pi^+ \pi^- 2\pi^0)=(4.0^{+1.1}_{-1.2})\%
\cdot (15.25 \pm 0.35)\%/(49.2 \pm 0.6)\%$, resulting in $\b(D_s^+ \to 2\pi^+
\pi^- 2\pi^0) = (1.24 \pm 0.36)\%$.

(3) The state $3\pi^+ 2\pi^- \pi^0$ receives a contribution from $D_s \to \phi
2\pi^+ \pi^-$, resulting in $\b(D_s^+ \to 3\pi^+ 2\pi^- \pi^0) = (0.18 \pm
0.03)\%$.

(4) The state $2\pi^+ \pi^- 3\pi^0$ receives a contribution from $D_s \to \phi
\pi^+ 2\pi^0$, resulting in $\b(D_s^+ \to2\pi^+ \pi^- 3\pi^0) = (0.12 \pm 0.03)\%$.
\bigskip

\noindent
{\bf F. Avoiding double counting from $D_s \to (\eta,\eta')+(\pi^+,\rho^+)$}
\medskip

The decays $D_s \to (\eta \pi^+,\eta'\pi^+,\eta\rho^+,\eta'\rho^+)$ can
populate numerous multi-pion final states.  Of these, the only inclusive
branching fraction explicitly quoted in Ref.\ \cite{PDG} is ${\cal B}(D_s \to
3 \pi^+ 2 \pi^- \pi^0) = (4.9 \pm 3.2)\%$.  We estimate using quoted branching
fractions that $\b(D_s \to \eta' \pi^+) \b(\eta' \to \eta \pi^+ \pi^-)
\b(\eta \to \pi^+ \pi^- \pi^0)=(0.39 \pm 0.04)\%$, leaving $(4.5\pm3.2)\%$
to be accounted for elsewhere.  It is this figure that we include in the
table of $D_s$ branching fractions.
\bigskip

\centerline{\bf V.  PROVISIONAL TABLES OF BRANCHING FRACTIONS}
\bigskip

We summarize $D_s$ branching fractions to leptonic and semileptonic final
states in Table \ref{tab:lbrs} and to hadronic final states in Table
\ref{tab:hbrs}.  Averages in the literature \cite{PDG} have been supplemented
where necessary with statistical model estimates, but an attempt has been
made to use as much input from data as possible.  We treat $\eta X$, $\eta' X$,
and $\omega X$ final states explicitly, but quote only final states for $\phi
X$.  The sum of the entries in
Tables \ref{tab:lbrs} and \ref{tab:hbrs} is $(102.7 \pm 5.3)\%$. 
The largest source of error is a rather early measurement~\cite{Barlag:1992ww},
$\b(D_s\to 3\pi^+ 2\pi^-\pi^0)=(4.9\pm 3.2)\%$, which includes a small
contribution from $D_s\to\eta'\pi^+$. Replacement of this number and
many estimated numbers with observed ones will be possible given the excellent
particle identification and electromagnetic calorimetry of the CLEO-c detector.
We do not extrapolate to other $6 \pi$ modes from $3\pi^+ 2\pi^- \pi^0$,
preferring to wait until it is better-measured.

The last 11 entries in Table \ref{tab:hbrs} sum to a branching fraction of
$(4.41 \pm 0.78)\%$ for hadronic Cabibbo-suppressed decays of $D_s$,
consistent with a fraction $|V_{cd}/V_{cs}|^2$ of the Cabibbo-favored hadronic
decays which amounts to $(4.24 \pm 0.27)\%$.
\bigskip

\centerline{\bf VI.  PREDICTED INCLUSIVE BRANCHING FRACTIONS}
\bigskip

With the branching fractions in Tables \ref{tab:lbrs} and \ref{tab:hbrs}, it
now becomes possible to calculate inclusive particle yields.  These are
summarized for pions in Table \ref{tab:inp}, for kaons in Table \ref{tab:ink},
and for $\eta,~\eta',~\phi$, and $\omega$ in Table \ref{tab:ine}.
Errors in $\pi^+$ and $\pi^-$ inclusive branching ratios can be reduced 
considerably by improving the branching fraction measurement for
$D_s\to 3\pi^+2\pi^-\pi^0$.

\begin{table}
\caption{$D_s$ branching fractions to leptonic and semileptonic modes
\cite{PDG,newtau,Yelton:2009cm}.  Only modes contributing to inclusive hadron
production are shown.  Modes with $\b < 10^{-3}$ are omitted.  Each $D_s \to X
\ell^+ \nu_\ell$ represents the assumed sum of $X e^+ \nu$ and $X \mu^+
\nu$ modes and is doubled from the value quoted in Ref.\ \cite{PDG}.
\label{tab:lbrs}}
\begin{center}
\begin{tabular}{c c c} \hline \hline
Mode $f$ & $\b(D_s \to f)$ (\%) & Remarks \\ \hline
$\tau^+ \nu_\tau$ & 6.6$\pm$0.6 & See Table \ref{tab:tau} for $\tau^+$ hadronic decays\\
$\eta \ell^+ \nu_\ell$ & 5.8$\pm$1.2 & See Table \ref{tab:eta} for $\eta$
hadronic decays \\
$\eta' \ell^+ \nu_\ell$ & 2.04$\pm$0.66 & See Table \ref{tab:etap} for $\eta'$
hadronic decays \\
$\phi \ell^+ \nu_\ell$ & 4.72$\pm$0.52 & See Table \ref{tab:phi} for $\phi$
hadronic decays \\ \hline
Total & 19.16$\pm$1.58 & \\ \hline \hline
\end{tabular}
\end{center}
\end{table}

\begin{table}
\caption{$D_s$ branching fractions to hadronic modes \cite{PDG}.  Values for
bracketed modes are inferred from arguments in Sections II and III.  Second errors are
systematic uncertainties obtained by differences between predictions of the
statistical isospin model and the model based on 
$q\bar q$ production. Modes with $\b < 10^{-3}$ are omitted.
\label{tab:hbrs}}
\begin{center}
\begin{tabular}{c c c} \hline \hline
Mode $f$ & $\b(D_s \to f)$ (\%) & Remarks \\ \hline
$[K^+ \ok]$ & 2.98$\pm$0.18 & Doubled from quoted $K^+ K_S$ value \\
$K^+ K^- \pi^+$ & 5.50$\pm$0.28 & \\
$[K^0 \ok \pi^+]$ & 5.76$\pm$0.96 & Quasi-2-body +  model for NR\\
$[K^+ \ok \pi^0]$ & 3.56$\pm$0.67 & " " \\
$K^+ K^- \pi^+ \pi^0$ & 5.6$\pm$0.5 & \\
$[K^0 \ok \pi^+ \pi^0]$ &4.36$\pm$1.68 & Quasi-2-body + model for NR \\
$[K^+ \ok \pi^+ \pi^-]$ & 1.92$\pm$0.26 &Doubled from quoted value with $K_S$\\
$[K^0 K^- \pi^+ \pi^+]$ & 3.28$\pm$0.24 & " "\\
$[K^+ \ok \pi^0 \pi^0]$ &0.80$\pm$0.67 & Quasi-2-body + model for NR \\
$K^+ K^- \pi^+ \pi^+ \pi^-$ & 0.88$\pm$0.16 & \\
$[K^0 \ok \pi^+ \pi^+ \pi^-]$&0.70$\pm$0.21 &$\phi(3\pi)$ + model\\
$[K^+ \ok \pi^+ \pi^- \pi^0]$ & 0.46$\pm$0.34 & " " \\
$[K^0 K^- \pi^+ \pi^+ \pi^0]$ & 0.23$\pm$0.16 & " " \\
$[K^+ K^- \pi^+ \pi^0 \pi^0]$ & 0.59$\pm$0.16 & " " \\
$[K^0 \ok \pi^+ \pi^0 \pi^0]$ & 0.47$\pm$0.15 & " " \\
$[K^+ \ok \pi^0 \pi^0 \pi^0]$ & 0.06$\pm$0.06 & " " \\
$\pi^+ \pi^+ \pi^-$ & 1.11$\pm$0.08 & \\
$[\pi^+ \pi^0 \pi^0]$ & 0.74$\pm$0.05$\pm 0.09$ & Model from $\pi^+ \pi^+
 \pi^-$ \\
$[\pi^+ \pi^+ \pi^- \pi^0]$ & 0.67$\pm$0.06 & From $\phi \pi^+$ \\
$\eta \pi^+$ & 1.58$\pm$0.21 & \\
$\omega \pi^+$ & 0.25$\pm$0.09 \\
$3 \pi^+ 2 \pi^-$ & 0.80$\pm$ 0.09 & \\
$[2 \pi^+ \pi^- 2\pi^0]$ & 3.00$\pm$0.41$\pm$0.24 & From $\phi \rho^+$ and 
model from $3 \pi^+ 2 \pi^-$ \\
$[\pi^+ 4 \pi^0]$ & 0.24$\pm$0.03$\pm$0.01 & Model from $3 \pi^+ 2 \pi^-$ \\
$\eta \rho^+$ & 13.0$\pm$2.2 & \\
$[3 \pi^+ 2 \pi^- \pi^0]$ & 4.5$\pm$3.2&Subtracting 0.39\% for $\eta' \pi^+$\\
$2 \pi^+ \pi^- 3 \pi^0$ & 0.12$\pm$0.03&From $D_s \to \phi \pi^+ \pi^0 \pi^0$\\
$\eta' \pi^+$ & 3.8$\pm$0.4 & \\
$\eta' \rho^+$ & 12.2$\pm$2.0 & \\
$[K^0 \pi^+]$ & 0.25$\pm$0.03 & Doubled from $K_S \pi^+$ \\
$K^+ \pi^0$ & 0.08$\pm 0.02$ & (Statistical: 0.25$\pm$0.03, 
$q\bar q$: unrestricted)\\
$K^+ \eta$ & 0.141$\pm$0.031 & \\
$K^+ \eta'$ & 0.16$\pm$0.05 & \\
$K^+ \pi^+ \pi^-$ & 0.69$\pm$0.05 & \\
$[K^0 \pi^+ \pi^0]$ & 0.61$\pm$0.04$^{+0.43}_{-0.26}$ & 
Model from $K^+ \pi^+ \pi^-$\\
$[K^+\pi^0\pi^0]$ & 0.23$\pm$0.02$^{+0.12}_{-0.23}$ & " " \\
$[K^0 \pi^+ \pi^+ \pi^-]$ &0.60$\pm$0.22& Doubled from $K_S\pi^+\pi^+\pi^-$ \\
$[K^+ \pi^+ \pi^- \pi^0]$ & 1.05$\pm$0.39$\pm$0.45 & Model from
 $K^0 \pi^+ \pi^+ \pi^-$ \\
$[K^0 \pi^+ \pi^0 \pi^0]$ & 0.45$\pm$0.17 & " " \\
$[K^+ 3 \pi^0]$ & 0.15$\pm$0.06$^{+0.08}_{-0.15}$& " " \\ \hline
Total & 83.57$\pm$5.05 & \\ \hline \hline
\end{tabular}
\end{center}
\end{table}

\begin{table}
\caption{Inclusive yields of pions from various final states in $D_s$ decays.
\label{tab:inp}}
\begin{center}
\begin{tabular}{c c c c c} \hline \hline
Mode & $\b(\%)$ & $\b(\pi^+)(\%)$ & $\b(\pi^0)(\%)$ & $\b(\pi^-)(\%)$ \\ \hline
$\tau^+ \nu_\tau$ & $6.6\pm0.6$ & 5.11$\pm$0.46 & 3.58$\pm$0.33 &
 0.98$\pm$0.09 \\
$\eta \ell^+ \nu_\ell$ & $5.8\pm1.2$ & 1.59$\pm$0.33 & 6.98$\pm$1.45 &
 1.59$\pm$0.33 \\
$\eta' \ell^+ \nu_\ell$ & 2.04$\pm$0.66 & 1.93$\pm$0.63 & 2.51$\pm$0.82 &
 1.93$\pm$0.63 \\
$\phi \ell^+ \nu_\ell$ & 4.72$\pm$0.52 & 0.74$\pm$0.08 & 0.79$\pm$0.09 &
 0.74$\pm$0.08 \\
$K^+ K^- \pi^+$ & 5.50$\pm$0.28 & 5.50$\pm$0.28 & 0 & 0 \\
$K^0 \ok \pi^+$ & 5.76$\pm$0.96 & 5.76$\pm$0.96 & 0 & 0 \\
$K^+ \ok \pi^0$ & 3.56$\pm$0.67 & 0 & 3.56$\pm$0.67 & 0 \\
$K^+ K^- \pi^+ \pi^0$ & $5.6\pm0.5$ & $5.6\pm0.5$ & $5.6\pm0.5$ & 0 \\
$K^0 \ok \pi^+ \pi^0$ & 4.36$\pm$1.68 & 4.36$\pm$1.68 & 4.36$\pm$1.68 & 0 \\
$K^+ \ok \pi^+ \pi^-$ & 1.92$\pm$0.26 & 1.92$\pm$0.26 & 0 & 1.92$\pm$0.26 \\
$K^0 K^- \pi^+ \pi^+$ & 3.28$\pm$0.24 & 6.56$\pm$0.48 & 0 & 0 \\
$K^+ \ok \pi^0 \pi^0$ & 0.80$\pm$0.67 & 0 & 1.60$\pm$1.34 & 0 \\
$K^+ K^- \pi^+ \pi^+ \pi^-$ & 0.88$\pm$0.16 & 1.76$\pm$0.32 & 0 &
 0.88$\pm$0.16\\
$K^0 \ok \pi^+ \pi^+ \pi^-$ & 0.70$\pm$0.21 & 1.40$\pm$0.42 & 0 &
 0.70$\pm$0.21 \\
$K^+ \ok \pi^+ \pi^- \pi^0$ & 0.46$\pm$0.34 & 0.46$\pm$0.34 & 0.46$\pm$0.34 &
 0.46$\pm$0.34 \\
$K^0 K^- \pi^+ \pi^+ \pi^0$ & 0.23$\pm$0.16 & 0.46$\pm$0.32 & 0.23$\pm$0.16 &
 0 \\
$K^+ K^- \pi^+ \pi^0 \pi^0$ & 0.59$\pm$0.16 & 0.59$\pm$0.16 & 1.18$\pm$0.32 &
 0 \\
$K^0 \ok \pi^+ \pi^0 \pi^0$ & 0.47$\pm$0.15 & 0.47$\pm$0.15 & 0.94$\pm$0.30 &
 0 \\
$K^+ \ok \pi^0 \pi^0 \pi^0$ & 0.06$\pm$0.06 & 0 & 0.18$\pm$0.18 & 0 \\
$\pi^+ \pi^+ \pi^-$ & 1.11$\pm$0.08 & 2.22$\pm$0.16 & 0 & 1.11$\pm$0.08 \\
$\pi^+ \pi^0 \pi^0$ & 0.74$\pm$0.10 & 0.74$\pm$0.10 & 1.48$\pm$0.20 & 0 \\
$\pi^+ \pi^+ \pi^- \pi^0$ & 0.67$\pm$0.06 & 1.34$\pm$0.12 & 0.67$\pm$0.06 &
 0.67$\pm$0.06 \\
$\eta \pi^+$ & 1.58$\pm$0.21 & 2.01$\pm$0.27 & 1.90$\pm$0.25 & 0.43$\pm$0.06 \\
$\omega \pi^+$ & 0.25$\pm$0.09 & 0.48$\pm$0.17 & 0.25$\pm$0.09 &
 0.22$\pm$0.08 \\
$3 \pi^+ 2 \pi^-$ & 0.80$\pm$ 0.09 & 2.40$\pm$0.27 & 0 & 1.60$\pm$0.18 \\
$2 \pi^+ \pi^- 2\pi^0$ & 3.00$\pm$0.48 & 6.00$\pm$0.96 & 6.00$\pm$0.96 &
 3.00$\pm$0.48 \\
$\pi^+ 4 \pi^0$ & 0.24$\pm$0.03 & 0.24$\pm$0.03 & 0.96$\pm$0.12 & 0 \\
$\eta \rho^+$ & 13.0$\pm$2.2 & 16.55$\pm$2.80 & 28.65$\pm$4.85 & 
 3.55$\pm$0.60 \\
$3 \pi^+ 2 \pi^- \pi^0$ & 4.5$\pm$3.2 & 13.5$\pm$9.6 & 4.5$\pm$3.2 &
9.0$\pm$6.4 \\
$2 \pi^+ \pi^- 3 \pi^0$ & 0.12$\pm$0.03 & 0.24$\pm$0.06 & 0.36$\pm$0.09 &
 0.12$\pm$0.03 \\
$\eta' \pi^+$ & 3.8$\pm$0.4 & 7.39$\pm$0.79 & 4.67$\pm$0.49 & 3.59$\pm$0.38 \\
$\eta' \rho^+$ & 12.2$\pm$2.0 & 23.74$\pm$3.89 & 27.21$\pm$4.46 &
 11.54$\pm$1.89 \\
$K^0 \pi^+$ & 0.25$\pm$0.03 & 0.25$\pm$0.03 & 0 & 0\\
$K^+ \pi^0$ & 0.08$\pm 0.02$ & 0 & 0.08$\pm 0.02$ & 0 \\
$K^+ \eta$ & 0.141$\pm$0.031 & 0.04$\pm$0.01 & 0.17$\pm$0.04 & 0.04$\pm$0.01 \\
$K^+ \eta'$ & 0.16$\pm$0.05 & 0.15$\pm$0.05 & 0.20$\pm$0.06 & 0.15$\pm$0.05 \\
$K^+ \pi^+ \pi^-$ & 0.69$\pm$0.05 & 0.69$\pm$0.05 & 0 & 0.69$\pm$0.05 \\
$K^0 \pi^+ \pi^0$ & 0.61$\pm$0.35 & 0.61$\pm$0.35 & 0.61$\pm$0.35 & 0\\
$K^+\pi^0\pi^0$ & 0.23$\pm$0.18 & 0 & 0.46$\pm$0.36 & 0 \\
$K^0 \pi^+ \pi^+ \pi^-$ & 0.60$\pm$0.22 & 1.20$\pm$0.44 & 0 & 0.60$\pm$0.22 \\
$K^+ \pi^+ \pi^- \pi^0$ & 1.05$\pm$0.60 & 1.05$\pm$0.60 & 1.05$\pm$0.60 &
 1.05$\pm$0.60 \\
$K^0 \pi^+ \pi^0 \pi^0$ & 0.45$\pm$0.17 & 0.45$\pm$0.17 & 0.90$\pm$0.34 & 0 \\
$K^+ 3 \pi^0$ & 0.15$\pm$0.13 & 0 & 0.45$\pm$0.39 & 0 \\ \hline
Total & & 125.5$\pm$11.1 & 112.5$\pm$8.0 & 46.6$\pm$6.8 \\ \hline \hline
\end{tabular}
\end{center}
\end{table}

\begin{table}
\caption{Inclusive yields of kaons from various final states in $D_s$ decays.
\label{tab:ink}}
\begin{center}
\begin{tabular}{c c c c c c} \hline \hline
Mode & $\b(\%)$ & $\b(K^+)(\%)$ & $\b(K^0)(\%)$ & $\b(K^-)(\%)$ & $\b(\ok)(\%)$
\\ \hline
$\tau^+ \nu_\tau$ & $6.6\pm0.6$ & 0.14$\pm$0.01 & 0.10$\pm$0.01 &0.03& 0.01 \\
$\phi \ell^+ \nu_\ell$ & 4.72$\pm$0.52 & 2.32$\pm$0.26 & 1.60$\pm$0.18 &
 2.32$\pm$0.26 & 1.60$\pm$0.18 \\
$K^+ \ok$ & 2.98$\pm$0.18 & 2.98$\pm$0.18 & 0 & 0 & 2.98$\pm$0.18 \\
$K^+ K^- \pi^+$ & 5.50$\pm$0.28 & 5.50$\pm$0.28 & 0 & 5.50$\pm$0.28 & 0 \\
$K^0 \ok \pi^+$ & 5.76$\pm$0.96 & 0 & 5.76$\pm$0.96 & 0 & 5.76$\pm$0.96 \\
$K^+ \ok \pi^0$ & 3.56$\pm$0.67 & 3.56$\pm$0.67 & 0 & 0 & 3.56$\pm$0.67 \\
$K^+ K^- \pi^+ \pi^0$ & $5.6\pm0.5$ & $5.6\pm0.5$ & 0 & $5.6\pm0.5$ & 0 \\
$K^0 \ok \pi^+ \pi^0$ & 4.36$\pm$1.68 & 0 & 4.36$\pm$1.68 & 0 & 4.36$\pm$1.68\\
$K^+ \ok \pi^+ \pi^-$ & 1.92$\pm$0.26 & 1.92$\pm$0.26 & 0 & 0 & 1.92$\pm$0.26\\
$K^0 K^- \pi^+ \pi^+$ & 3.28$\pm$0.24 & 0 & 3.28$\pm$0.24 & 3.28$\pm$0.24 & 0\\
$K^+ \ok \pi^0 \pi^0$ & 0.80$\pm$0.67 & 0.80$\pm$0.67& 0 & 0 & 0.80$\pm$0.67\\
$K^+ K^- \pi^+ \pi^+ \pi^-$ & 0.88$\pm$0.16 & 0.88$\pm$0.16 & 0 &
 0.88$\pm$0.16 & 0 \\
$K^0 \ok \pi^+ \pi^+ \pi^-$ & 0.70$\pm$0.21 & 0 & 0.70$\pm$0.21 & 0 &
 0.70$\pm$0.21 \\
$K^+ \ok \pi^+ \pi^- \pi^0$ & 0.46$\pm$0.34 & 0.46$\pm$0.34 & 0 & 0 &
 0.46$\pm$0.34 \\
$K^0 K^- \pi^+ \pi^+ \pi^0$ & 0.23$\pm$0.16 & 0 & 0.23$\pm$0.16 &
 0.23$\pm$0.16 & 0 \\
$K^+ K^- \pi^+ \pi^0 \pi^0$ & 0.59$\pm$0.16 & 0.59$\pm$0.16 & 0 &
 0.59$\pm$0.16 & 0 \\
$K^0 \ok \pi^+ \pi^0 \pi^0$ & 0.47$\pm$0.15 & 0 & 0.47$\pm$0.15 & 0 &
 0.47$\pm$0.15 \\
$K^+ \ok \pi^0 \pi^0 \pi^0$ & 0.06$\pm$0.06 & 0.06$\pm$0.06 & 0 & 0 &
 0.06$\pm$0.06 \\
$K^0 \pi^+$ & 0.25$\pm$0.03 & 0 & 0.25$\pm$0.03 & 0 & 0 \\
$K^+ \pi^0$ & 0.08$\pm 0.02$ & 0.08$\pm 0.02$ & 0 & 0 & 0 \\
$K^+ \eta$ & 0.141$\pm$0.031 & 0.14$\pm$0.03 & 0 & 0 & 0 \\
$K^+ \eta'$ & 0.16$\pm$0.05 & 0.16$\pm$0.05 & 0 & 0 & 0 \\
$K^+ \pi^+ \pi^-$ & 0.69$\pm$0.05 & 0.69$\pm$0.05 & 0 & 0 & 0 \\
$K^0 \pi^+ \pi^0$ & 0.61$\pm$0.35 & 0 & 0.61$\pm$0.35 & 0 & 0 \\
$K^+\pi^0\pi^0$ & 0.23$\pm$0.18 & 0.23$\pm$0.18 & 0 & 0 & 0 \\
$K^0 \pi^+ \pi^+ \pi^-$ & 0.60$\pm$0.22 & 0 & 0.60$\pm$0.22 & 0 & 0 \\
$K^+ \pi^+ \pi^- \pi^0$ & 1.05$\pm$0.60 & 1.05$\pm$0.60 & 0 & 0 & 0 \\
$K^0 \pi^+ \pi^0 \pi^0$ & 0.45$\pm$0.17 & 0 & 0.45$\pm$0.17 & 0 & 0 \\
$K^+ 3 \pi^0$ & 0.15$\pm$0.13 & 0.15$\pm$0.13 & 0 & 0 & 0 \\ \hline
Total & & 27.3$\pm$1.4 & 18.4$\pm$2.0 & 18.4$\pm$0.7 & 22.7$\pm$ 2.2 \\
\hline \hline
\end{tabular}
\end{center}
\end{table}

\begin{table}
\caption{Inclusive yields of $\eta$, $\eta'$, $\phi$, and $\omega$ from various
final states in $D_s$ decays.
\label{tab:ine}}
\begin{center}
\begin{tabular}{c c c c c c} \hline \hline
Mode & $\b(\%)$ & $\b(\eta)(\%)$ & $\b(\eta')(\%)$ & $\b(\phi)(\%)$ &
 $\b(\omega)(\%)$
\\ \hline
$\eta \ell^+ \nu_\ell$ & 5.8$\pm$1.2 & 5.8$\pm$1.2 & 0 & 0 & 0 \\
$\eta' \ell^+ \nu_\ell$ & 2.04$\pm$0.66 & 1.33$\pm$0.43 & 2.04$\pm$0.66 & 0 &
0.06$\pm$0.02 \\
$\phi \ell^+ \nu_\ell$ & 4.72$\pm$0.52 & 0.06$\pm$0.01 & 0 & 4.72$\pm$0.52 &0\\
$\phi \pi^+$ & 4.38$\pm$0.35 & 0.06$\pm$0.01 & 0 & 4.38$\pm$0.35 & 0 \\
$\phi \rho^+$ & 8.13$\pm$2.34 & 0.11$\pm$0.03 & 0 & 8.13$\pm$2.34 & 0 \\
$\phi 2\pi^+ \pi^-$ & 1.20$\pm$0.22 & 0.02 & 0 & 1.20$\pm$0.22 & 0 \\
$\phi \pi^+ 2 \pi^0$ & 0.80$\pm$0.15 & 0.01 & 0 & 0.80$\pm$0.15 & 0 \\
$\eta \pi^+$ & 1.58$\pm$0.21 & 1.58$\pm$0.21 & 0 & 0 & 0 \\
$\omega \pi^+$ & 0.25$\pm$0.09 & 0 & 0 & 0 & 0.25$\pm$0.09 \\
$\eta \rho^+$ & 13.0$\pm$2.2 & 13.0$\pm$2.2 & 0 & 0 & 0 \\
$\eta' \pi^+$ & 3.8$\pm$0.4 & 2.48$\pm$0.26 & 3.8$\pm$0.4 & 0 & 0.11$\pm$0.01\\
$\eta' \rho^+$ & 12.2$\pm$2.0 & 7.97$\pm$1.32 & 12.2$\pm$2.0 & 0 &
 0.37$\pm$0.07 \\
$K^+ \eta$ & 0.141$\pm$0.031 & 0.14$\pm$0.03 & 0 & 0 & 0 \\
$K^+ \eta'$ & 0.16$\pm$0.05 & 0.10$\pm$0.03 & 0.16$\pm$0.05 & 0 & 0 \\
\hline
Total & & 32.7$\pm$2.9 & 18.2$\pm$2.1 & 19.2$\pm$2.4 & 0.8$\pm$0.1 \\
\hline \hline
\end{tabular}
\end{center}
\end{table}

\bigskip
\centerline{\bf VII.  CONCLUSIONS}
\bigskip

We have calculated the inclusive branching fractions of $D_s$ mesons to several
species, using the fact that the observed branching fractions, together with
modest assumptions about unseen charge states, account for all the $D_s$ decays
to an accuracy of about 5\%.  Calculations of branching ratios for unseen
modes, based mostly on a statistical isospin model, involve small systematic
theoretical uncertainties estimated by comparison with a model using
quark-antiquark pair production.  While many aspects of this analysis bear some
resemblance to an itemized tax return, several notable features have emerged.

\begin{itemize}

\item The greatest errors on extracting inclusive branching fractions from
exclusive modes are due to a few final states, notably $3 \pi^+ 2 \pi^- \pi^0$,
$\eta \rho^+$, and $\eta' \rho^+$.  Improvement of information on these modes
would be very helpful.

\item The large predicted values for 
the inclusive branching ratios
$\b(\eta)$ and $\b(\eta')$ may be helpful
in determining whether, as suspected in SU(3) fits (see, e.g., Ref.\
\cite{Bhattacharya:2008ke}) and in factorization calculations or modifications
\cite{Jessop:1998bc,fact} based on the observed semileptonic decays
$D_s \to \eta^{(\prime)} \ell \nu$, the branching
fractions for $D_s \to \eta \rho^+$ and particularly $\eta' \rho^+$ are too
high.  It could be possible that part of the large ``signals'' for $D_s \to
\eta^{(\prime)} \rho^+$ come from misidentified kinematic reflections from
other final states.  

For example, both $\eta' (\to \pi^+ \pi^- \eta) \rho^+
(\to \pi^+ \pi^0)$ and 
$\eta \pi^+ \omega (\to \pi^+ \pi^- \pi^0)$ contain
the same final-state particles $\eta 2 \pi^+ \pi^- \pi^0$.  In the CLEO paper
reporting large branching fractions for $D_s \to \eta^{(\prime)} \rho^+$
(where $\eta$ was identified by its two-photon decay mode) 
\cite{Jessop:1998bc}, no distributions in $M(\eta)$ or $M(\eta')$ were shown.
If a sizable fraction of the measured value $\b(D_s\to\eta'\rho^+)=(12.2\pm 2.0)\%$ 
is, for instance, due to underlying $D_s\to \eta\pi^+\omega$ events, then our prediction 
for inclusive  $\b(\eta')$ becomes smaller while $\b(\omega)$ becomes 
correspondingly larger. The effect of such events on $\b(\eta)$ combines a decrease 
from a smaller value of $\b(D_s\to\eta'\rho^+)$, where $\eta'$ decays dominantly to states 
involving $\eta$ (see Table VI), and a positive contribution from $\b(D_s\to\eta\pi^+\omega)$.

Similarly, the decay mode $D_s\to \omega(\to \pi^0\gamma)\pi^+\pi^0$, where one
photon from $\pi^0$ may be missing in some events, can mimic 
$D_s \to \eta(\to \gamma\gamma)
\rho^+(\to\pi^+\pi^0)$. If a small fraction of $\b(D_s\to\eta\rho^+)=(13.0\pm 2.2)\%$ 
stems from underlying $D_s\to\omega\pi^+\pi^0$ events, then our predictions for 
inclusive $\b(\eta)$ and  $\b(\omega)$ decrease and increase, respectively, roughly 
(i.e. neglecting differences in efficiencies) 
by a ratio $\b(\omega\to \pi^0\gamma):\b(\eta\to\gamma\gamma) = 1:4.4$.

\item The measurement of $\b(\eta)$, $\b(\eta')$, and $\b(\omega)$ may help
to shed light on specific decay mechanisms.  For example, the decay $D_s \to
\omega \pi^+ \pi^0$, represented by the quark annihilation process $c\bar s
\to u\bar d$, could have a sizable branching ratio.  While $D_s\to \omega\pi^+$
with $\b=(0.25\pm 0.09)\%$ is forbidden by G-parity and requires preradiation
of $\omega$ or rescattering~\cite{Gronau:2009mp}, $D_s \to \omega \pi^+ \pi^0$
can occur through ordinary quark annihilation (as long as some mechanism
overcomes helicity suppression), and is expected to have a considerably larger
decay rate.  As a second example, the decay $D_s \to \omega \pi^+ \eta$ could
arise either from WA or from the transition $c \bar s \to s \bar s +$ (charged
weak vector current), where the charged weak vector current produces $\omega
\pi^+$.

\end{itemize}

We look forward to experimental tests of these predictions.

\newpage
\centerline{\bf ACKNOWLEDGMENTS} 
\bigskip 

We thank Ed Thorndike and Fan Yang for useful discussions.
J. L. R. would like to thank the Physics Deparment of the Technion for its
hospitality during part of this study.  This work was supported in part by 
the United States Department of Energy through Grant No.\ DE-FG02-90ER-40560. 
\bigskip

{\it Note added in proof:}~  Inclusive $D_s$ decays have now been reported by
the CLEO Collaboration~\cite{CLEO-Ds}.  They are largely in agreement with the
predictions of Tables XVII -XIX, with the notable exceptions of $\b(\eta') =
(11.7 \pm 1.7 \pm 0.7)\%$ and $\b(\omega) = (6.1 \pm 1.4 \pm 0.3)\%$.
Thus, current world averages \cite{PDG} apparently overestimate the
sum of $D_s$ branching fractions involving $\eta'$ and greatly
underestimate those involving $\omega$.

\end{document}